\documentclass[12pt]{article}

\usepackage{amsfonts}
\usepackage{epsfig}
\usepackage{latexsym}
\usepackage{graphicx}
\def\bequ{\begin{equation}}
\def\eequ{\end{equation}}
\def\barr{\begin{array}}
\def\earr{\end{array}}
\def\half{{1\over 2}}
\def\ben{\begin{equation}}
\def\een{\end{equation}}
\def\bena{\begin{eqnarray}}
\def\eena{\end{eqnarray}}

\renewcommand{\theequation}{\arabic{section}.\arabic{equation}}
\def\spa#1{\phantom{\fbox{\rule[-#1cm]{0cm}{0cm}}}}

\def\b1{e^0}

\newcommand{\be}{\begin{equation}}
\newcommand{\ee}{\end{equation}}
\def\bea{\begin{eqnarray}}
\def\eea{\end{eqnarray}}

\def\del {\partial}
\def\nn{\nonumber}

\def\half {{1 \over 2}}
\def\be{\begin{equation}}
\def\ee{\end{equation}}
\def\bea{\begin{eqnarray}}
\def\eea{\end{eqnarray}}

\def\lesssim{\mathrel{\hbox{\rlap{\hbox{\lower4pt\hbox{$\sim$}}}\hbox{$<$}}}}
\def\gtrsim{\mathrel{\hbox{\rlap{\hbox{\lower4pt\hbox{$\sim$}}}\hbox{$>$}}}}

\begin{document}

\hfuzz=100pt
\title{{\Large \bf{Energy Quantisation in Bulk Bouncing Tachyon}}}
\author{\\Shinji Hirano\footnote{hirano@physics.technion.ac.il}
  \spa{0.5} \\
{{\it Department of Physics, Technion}},
\\ {{\it Israel Institute of Technology}},
\\ {{\it Haifa 32000, Israel}}}

\date{February, 2005}

\maketitle
\centerline{}

\begin{abstract}
We argue that the closed string energy in the bulk bouncing tachyon 
background is to be quantised in a simple manner 
as if strings were trapped in a finite time interval.
We discuss it from three different viewpoints;  
(1) the timelike continuation of the sinh-Gordon model,
(2) the dual matrix model description of the ($1+1$)-dimensional
string theory with the bulk bouncing tachyon condensate, 
(3) the $c_L=1$ limit of the timelike Liouville theory 
with the dual Liouville potential turned on. 
There appears to be a parallel between the bulk bouncing tachyon 
and the full S-brane of D-brane decay. 
We find the critical value $\lambda_c$ of the bulk bouncing tachyon
coupling which is analogous to $\lambda_o=\half$ of the full S-brane
coupling, at which the system is thought to be at the bottom of the
tachyon potential. 
\end{abstract}

%%%%%%%%%%%%%%%%%%%%%%%%%%%%%%%%%%%%%%%%%%%%%%%%%%%%%%%%%%%%%%%%%%%%%
\section{Introduction}
%%%%%%%%%%%%%%%%%%%%%%%%%%%%%%%%%%%%%%%%%%%%%%%%%%%%%%%%%%%%%%%%%%%%

In recent years there has been a good amount of progress in
understanding the dynamics of tachyons localised in space 
-- the open string tachyon of D-brane decay 
\cite{Sen:2004nf}
and twisted closed string tachyon of orbifold decay
\cite{Adams:2001sv} (see also \cite{Headrick:2004hz} for a review).  

There is little understanding of what would happen when the bulk
tachyon condenses.
The bulk tachyon indicates the instability of the flat spacetime, 
so it is hard to imagine what kind of spacetime we would end up with
after the decay. 
A common speculation is that the spacetime disappears as the bulk
tachyon condenses. 
As closed strings propagate in the tachyon media in spacetime,
they would feel harder to go through it than in the spacetime 
without tachyon condensate.
If the endpoint of the tachyon decay is at $T_{blk}=\infty$, 
it would be too hard for strings to propagate at all
in the infinitely stiff tachyon media in spacetime.  
In fact it may be noted that the dual matrix model description
\cite{Minic:1991rk,Karczmarek:2003pv} of the bulk tachyon condensate
in ($1+1$)-dimensions seems to suggest this speculation to be true, 
even though in ($1+1$)-dimensions the closed string tachyon, 
being massless, does not indicate an instability and  
its condensation does not represent the decay process. 

A class of marginal deformations of conformal field theory (CFT) can 
describe the tachyon condensation with certain profiles. 
However, these deformations are not a small perturbation, and
in general it is not easy to solve the interacting CFTs with these
deformations. So finding such exactly solvable CFTs can
potentially lead to progress in understanding the tachyon dynamics.
It was indeed the case for the open string tachyon of D-brane decay. 
A nontrivial exact $c=1$ boundary CFT (BCFT) -- 
the massless limit of the boundary sine-Gordon model -- 
was solved in \cite{Callan:1994ub,Polchinski:1994my}. 
Later it was applied by means of a timelike continuation 
to the bouncing open string tachyon (full S-brane) of 
D-brane decay  by Sen \cite{Sen:2002nu,Sen:2002in}. 
Subsequently a similar BCFT was studied in
\cite{Gutperle:2003xf,Larsen:2002wc}
in its application to the rolling open string tachyon 
(half S-brane).\footnote{Following \cite{Lambert:2003zr}, 
we refer to the bouncing and rolling open
string tachyon as full S-brane and half S-brane respectively 
-- a coinage after S(pacelike)-brane of \cite{Gutperle:2002ai}.}    
The BCFT descriptions of the full and half S-branes
led to significant progress in understanding the D-brane decay.

In the bulk case, a nontrivial $c=1$ CFT was found by Runkel and Watts
\cite{Runkel:2001ng} as a limit of the minimal CFTs. It was later
proved by Schomerus and Fredenhagen 
\cite{Schomerus:2003vv,Fredenhagen:2004cj} 
that this CFT is indeed a $c=1$ limit of the Liouville CFT 
\cite{Dorn:1994xn, Zamolodchikov:1995aa}.
In the same spirit as Sen, 
it was then applied by means of a timelike continuation
to the bulk rolling tachyon 
by Strominger and Takayanagi \cite{Strominger:2003fn},
and Schomerus \cite{Schomerus:2003vv}. 
These developments can potentially be important steps towards better
understanding the bulk tachyon condensation.

In this paper we will consider the bulk bouncing tachyon
from three different viewpoints;
(1) the timelike continuation of the sinh-Gordon (ShG) model,
(2) the dual matrix model description of the ($1+1$)-dimensional
string theory with the bulk bouncing tachyon condensate, 
(3) the $c_L=1$ limit of the timelike Liouville theory 
with the dual Liouville potential turned on. 
We will study in particular the closed string spectrum in the bulk
bouncing tachyon background, 
and propose that the closed string energy is to be quantised as if
strings were trapped in a finite time interval.
 
The organisation of the paper is as follows. In section 2, we briefly 
remind the reader what the bulk bouncing tachyon is and its relation
to the Liouville theory.
In section 3, we study the timelike continuation of the ShG model and
argue that the energy is to be quantised, based on Zamoldchikovs'
argument of the momentum quantisation in the spacelike ShG model. 
In section 4, we study the dual matrix model of the
($1+1$)-dimensional string theory with the bulk bouncing tachyon
condensate. In particular, we discuss the spectrum of the theory
and find an agreement with the result in section 3.
In section 5, we make an attempt to construct the $c_L=1$ limit of the 
Liouville theory with the dual Liouville potential turned on.
We discuss how the restriction on the energy spectrum can arise  
and how the results in the previous sections could potentially be
consistent with this Liouville CFT.
We summarise our results in section 6 and close the section with brief
discussions.
In appendix A, we show some of the details of the calculation in
section 5. 

%%%%%%%%%%%%%%%%%%%%%%%%%%%%%%%%%%%%%%%%%%%%%%%%%%%%%%%%%%%%%%%%%%%%%%
\section{Bulk bouncing tachyon}
\setcounter{equation}{0}
%%%%%%%%%%%%%%%%%%%%%%%%%%%%%%%%%%%%%%%%%%%%%%%%%%%%%%%%%%%%%%%%%%%%%

We will consider the Liouville theory with the dual Liouville
potential turned on:
\begin{equation}
S={1 \over 4\pi}\int_{\Sigma}
d^2x\sqrt{g}\left(g^{ab}\del_a\phi\del_b\phi 
+QR_g\phi+4\pi\mu e^{2b\phi}
+4\pi\widetilde{\mu} e^{2\phi/b}\right)\ ,
\label{liouvilleaction}
\end{equation}
where $Q=b+1/b$ and the central charge of this theory is
$c_L=1+6Q^2$.\footnote{We will call $\mu$ cosmological constant (CC)
and $\widetilde{\mu}$ dual CC as commonly referred to as in
literatures.}

In particular, we are interested in its application to the bouncing
bulk tachyon. So we will eventually focus on the case $b=i$ together
with the Wick-rotation $\phi\to -i\phi$. 
This is a generalisation of the $c_L=1$ timelike Liouville theory
of Strominger and Takayanagi \cite{Strominger:2003fn}, and 
Schomerus \cite{Schomerus:2003vv}.
In this case, the potential $V(-i\phi)$ becomes hyperbolic cosine, 
as one can see it by shifting the
Wick-rotated Liouville field $\phi$ to 
$\phi+{1\over 4}\log(\widetilde{\mu}/\mu)$, which gives
\be
V_{b=i}(-i\phi)=sign(\mu)\times 2\lambda\cosh(2\phi)\ ,
\ee
where $\lambda=\sqrt{\mu\widetilde{\mu}}$ 
and we assume that $\mu\widetilde{\mu}>0$. 
Thus the $c_L=1$ theory we would like to consider is a massless limit
of the timelike sinh-Gordon (TShG) model. 

In application to string theory, this corresponds to having the bulk
tachyon profile to be hyperbolic cosine of the time coordinate $\phi$.
Classically the bulk tachyon $T_{blk}$ comes in 
from $T_{blk}=sign(\mu)\times\infty$ 
in the infinite past ($\phi=-\infty$), reaches at the value of  
$T_{blk}=sign(\mu)\lambda$ at $\phi=0$, and   
bounces back to $T_{blk}=sign(\mu)\times\infty$ 
in the infinite future ($\phi=+\infty$).

Without knowing the shape of the bulk tachyon potential, it may not
be clear what this tachyon profile really represents. 
However, it is tempting to interpret it as a process of
spacetime decay analogous to Sen's bouncing open string tachyon 
\cite{Sen:2002nu,Sen:2002in} of D-brane decay.
%where $T_{open}=\infty$ corresponds to the bottom of the tachyon
%potential.  
Thus in the critical string theory application, 
the bulk bouncing tachyon would be thought of as 
a process of homogeneous decay of the flat spacetime.
We note that the analogy can be seen more manifestly in the 
dual matrix model description of the ($1+1$)-dimensional string theory
with the bulk bouncing tachyon condensate 
\cite{Minic:1991rk,Karczmarek:2003pv}, as we will review in section 4.

The main purpose of our paper is to understand a kinematical
aspect -- the spectrum -- of the bulk bouncing tachyon background 
described by (\ref{liouvilleaction}).\footnote{The open string spectrum
in the boundary massless sine-Gordon model -- the spacelike
continuation-back of the full S-brane -- was computed in 
\cite{Polchinski:1994my,Gaberdiel:2001zq}.} 

%%%%%%%%%%%%%%%%%%%%%%%%%%%%%%%%%%%%%%%%%%%%%%%%%%%%%%%%%%%%%%%%%%%%%
\section{Timelike sinh-Gordon model}
\setcounter{equation}{0}
%%%%%%%%%%%%%%%%%%%%%%%%%%%%%%%%%%%%%%%%%%%%%%%%%%%%%%%%%%%%%%%%%%%%%

Since the bulk bouncing tachyon is described by a massless limit of
the TShG model, it would be a good starting 
point to look at the spacelike ShG model. 
On this score, we would like to put an emphasis on 
Zamolodchikovs' momentum quantisation in the ShG model
(see Sect.~6 of Ref. \cite{Zamolodchikov:1995aa} and also 
\cite{Ahn:1999un, Ahn:1999dz, Ahn:2000ki} for its
application to other integrable models).

Consider the ShG model on the cylinder of circumference
$\epsilon$: 
\be
S_{ShG}={1\over 4\pi}\int dt\int_0^{\epsilon} dx\left[
(\del_t\phi)^2-(\del_x\phi)^2
-8\pi\lambda_{ShG}\cosh(2b\phi)\right]\ ,
\ee
where $\phi(t,x)=\phi(t,x+\epsilon)$. This is a massive theory and the
coupling $\lambda_{ShG}(>0)$ has the dimension $\sim
[\mbox{mass}]^{2+2b^2}$.  
It is convenient to apply the scaling transformation,  
$(t,x)\to {2\pi\over \epsilon}(t,x)$. Since the scaling dimensions of $\phi$
and $\cosh(2b\phi)$ are $0$ and $-2b^2$ respectively, the
ShG action then becomes
\be
S_{ShG}={1\over 4\pi}\int dt\int_0^{2\pi} dx\left[
(\del_t\phi)^2-(\del_x\phi)^2
-8\pi\lambda_{ShG}\left({\epsilon\over 2\pi}\right)^{2+2b^2}
\cosh(2b\phi)\right]\ .
\ee
First of all, the ShG potential is a confining potential, and thus one
may expect to have the momentum to be quantised. 

Zamolodchikovs made it more precise;
We can regard the ShG model as the Liouville theory with the potential 
$\lambda_{ShG} e^{-2b\phi}$ on the left (in the $\phi$-coordinate) 
and $\lambda_{ShG} e^{2b\phi}$ on the right.
In fact the locations of the walls of the Liouville potentials are
given by  
$\phi_{left}={1\over 2b}
\ln\left(\lambda_{ShG}(\epsilon/2\pi)^{2+2b^2}\right)$ on the
left and 
$\phi_{right}=-{1\over 2b}
\ln\left(\lambda_{ShG}(\epsilon/2\pi)^{2+2b^2}\right)$ on the
right. 
As we go to the UV ($\epsilon\ll 1$), two walls are well
separated, and in the middle ($\phi_{left}\ll\phi\ll\phi_{right}$)  
the potential is free. So if we send a free wave from the middle, 
say, to the right wall,
it will be reflected as if in the Liouville theory with the potential 
$\lambda_{ShG}\phi^{2b\phi}$, 
then propagate through leftwards freely in the middle, 
be again reflected at the left wall 
this time as if in the Liouville theory with
the potential $\lambda_{ShG}\phi^{-2b\phi}$, 
and then come back to the middle.

The wavefunction after the first reflection is given by
\be
\Psi_P[\phi]\sim 
e^{2iP\phi}+(\epsilon/2\pi)^{-4iPQ}S_b(P)e^{-2iP\phi}\ ,
\ee
where $S_b(P)$ is the Liouville reflection amplitude
\be
S_b(P)=-\left(\pi\lambda_{ShG}\gamma(b^2)
\right)^{-2iP/b}
\frac{\Gamma(1+2iP/b)\Gamma(1+2iPb)}
{\Gamma(1-2iP/b)\Gamma(1-2iPb)}\ .
\label{reflection}
\ee
Then the wavefunction after the second reflection follows 
\be
\Psi_P[\phi]\sim 
(\epsilon/2\pi)^{-8iPQ}S_{-b}(-P)S_b(P)e^{2iP\phi}
+(\epsilon/2\pi)^{-4iPQ}S_b(P)e^{-2iP\phi}\ .
\ee
These two wavefunctions have to match up, which requires the consistency
condition
\be
(\epsilon/2\pi)^{-8iPQ}S_{-b}(-P)S_b(P)
=(\epsilon/2\pi)^{-8iPQ}S_b(P)^2=1\ .
\label{smomq}
\ee
This is the momentum quantisation.

Now we can repeat a similar analysis for the TShG model obtained by 
$b\to i\beta$ and $\phi\to -i\phi$,
\be
S_{TShG}={1\over 4\pi}\int dt\int_0^{2\pi} dx\left[
-(\del_t\phi)^2+(\del_x\phi)^2
-8\pi\lambda_{TShG}\left({\epsilon\over 2\pi}\right)^{2-2\beta^2}
\cosh(2\beta\phi)\right]\ ,
\label{TShG}
\ee
but with $\lambda_{TShG}<0$. 
In order to have the confining potential, $\lambda_{TShG}$ must be
negative, 
since the sign of the kinetic term is reversed for the timelike
theory. 

The energy is negative definite, and 
the action of the timelike theory, after the Wick-rotation of
the world-sheet time, is bounded from above. 
So we need to find, if possible, a good prescription to define this
timelike theory.  
Fortunately, the timelike Liouville theories which are ingredients of
the TShG model (\ref{TShG}) were constructed by 
Schomerus \cite{Schomerus:2003vv}, Strominger and
Takayanagi \cite{Strominger:2003fn}.
They assumed that $\lambda_{TShG}>0$, but there appears to be no
obstruction to continue $\lambda_{TShG}$ to negative value in their
results. 
Futhermore, the spacelike theory from which the timelike theory is
analytically continued has an oscillating potential. 
%So the sign of $\lambda_{TShG}$ does not seem to matter in defining
%the timelike theory. 

As far as the reflection amplitude is
concerned, it can be obtained in a straightforward manner, simply
replacing $b$ by $i\beta$ together with the analytic continuation
$P\to -i\omega$ in (\ref{reflection}). Then the argument which leads
to the momentum quantisation is exactly the same as in the spacelike
case. 
Thus we have   
\be
(\epsilon/2\pi)^{-8i\omega Q_{\beta}}S_{-i\beta}(i\omega)
S_{i\beta}(-i\omega)
=(\epsilon/2\pi)^{-8i\omega Q_{\beta}}S_{i\beta}(-i\omega)^2=1\ ,
\label{tshmom}
\ee
where $Q_{\beta}=\beta-1/\beta$. 
This is the energy quantisation we would like to claim.

In the massless limit $\beta\to 1$, this becomes\footnote{More precisely, 
as we will see in section 5, 
the $\lim_{\beta\to 1}S_{i\beta}(i\omega)$ and 
$\lim_{\beta\to 1}S_{-i\beta}(i\omega)$ are given by the first and
second equation in (\ref{singlelimit}) respectively, where 
$\omega_1=\omega_2=\omega\ ,\,\omega_3=0$ and 
$\lambda_{TShG}=\mu=\widetilde{\mu}$. Thus the expression
(\ref{TShGq}) is slightly heuristic.}
\be
\lim_{\beta\to 1}
\left(\pi\lambda_{TShG}\gamma(-\beta^2)\right)^{4i\omega}=1\ .
\label{TShGq}
\ee
Since the $\lim_{\beta\to 1}\gamma(-\beta^2)$ diverges, we need to
take the $\lambda_{TShG}\to 0$ limit such that the  
$\lim_{\beta\to 1}\left(\pi\lambda_{TShG}\gamma(-\beta^2)\right)$ 
is kept finite. 
We will give more precise prescription to take this limit below, 
but let us denote 
$\lim_{\beta\to 1}\left(\pi\lambda_{TShG}\gamma(-\beta^2)\right)$ as 
$(\lambda_{TShG})_{ren}$ for the moment.
Then the energy quantisation (\ref{TShGq}) reads
\be
\omega={\pi n\over 2\ln(\lambda_{TShG})_{ren}}\ ,
\qquad n\in\mathbb{Z}_+\ .
\label{eqTShG}
\ee
Since the positive and negative $\omega$ are related and thus not
independent due to the reflection, we have taken only the positive
integer $n$. The zero mode $n=0$ corresponds to a constant
wavefunction and so should be removed from a complete set of states in
the Hilbert space with the discrete spectrum.

In \cite{Fredenhagen:2003ut}, Fredenhagen and Schomerus studied the 
minisuperspace approximation of the rolling and bouncing tachyons.
In particular, in the bouncing tachyon case, they found the discrete
energy spectrum. They considered the case of $\lambda_{TShG}>0$.
In that case, the potential is not confining, but one still needs to
impose the boundary conditions at $\phi=\pm\infty$, since a classical
particle in the exponential potentials can reach at $\phi=\pm\infty$
in a finite time. 
As the system is then effectively in a finite box, the energy is
quantised. 
Our quantisation condition is different from theirs. This is not a
contradiction, since the boundary conditions at $\phi=\pm\infty$ 
would not yield the same effect as the confining potential. Also the
minisuperspace approximation is a good approximation in the
semi-classical limit $\beta\to 0$, while we are considering the
$\beta\to 1$ limit.

As long as the theory is massive, two walls are well separated in the
UV ($\epsilon\ll 1$), which validates the above argument in the UV. 
Moreover, in this case, there is an alternative check of the
momentum quantisation (\ref{smomq}) 
by the Thermodynamic Bethe Ansatz for the spacelike theory.

In the massless limit, the theory does not depend on the scale
$\epsilon$. So the picture of the well-separated walls is only true  
when $|(\lambda_{TShG})_{ren}|\ll 1$.
However, we would like to argue that the energy quantisation 
(\ref{eqTShG}) is valid for generic value of
$(\lambda_{TShG})_{ren}$. 
In the following sections, 
we will provide two alternative arguments which support our claim.

%%%%%%%%%%%%%%%%%%%%%%%%%%%%%%%%%%%%%%%%%%%%%%%%%%%%%%%%%%%%%%%%%%
\section{The matrix model}
\setcounter{equation}{0}
%%%%%%%%%%%%%%%%%%%%%%%%%%%%%%%%%%%%%%%%%%%%%%%%%%%%%%%%%%%%%%%%%%

We would like to use the matrix model of $(1+1)$-dimensional string
theory in order to learn the $c_L=1$ Liouville theory 
with the dual Liouville potential turned on.
This is an idea explored in \cite{Minic:1991rk} where the $c=1$ matrix
model was used to study the Liouville theory with $c_L<25$.

The ($1+1$)-dimensional string theory with the linear dilaton is dual
to the double scaled matrix quantum mechanics (MQM) 
\cite{Gross:1990ay,Brezin:1989ss,Ginsparg:1990as,Parisi:1990jy,Gross:1990ub} 
and \cite{McGreevy:2003kb,Klebanov:2003km}. 
The $U(N)$ MQM  is equivalent to the system of $N$ non-interacting
fermions trapped in the MQM potential (fermi liquid) 
\cite{Brezin:1977sv}.
In the double scaling limit 
\cite{Brezin:1989ss,Ginsparg:1990as,Parisi:1990jy,Gross:1990ub}, 
$N$ is taken to the infinity and the
fermi level $-\mu_{bare}$ as measured from the top of the potential 
is taken to zero, while keeping $N\mu_{bare}(=1/g_s)$ fixed finite.
Then the double scaled MQM is the second quantised free fermion theory
in the inverted harmonic oscillator potential:
\be
H_F=\half\int_{-\infty}^{\infty} dx\left(
\del_x\Psi^{\dagger}\del_x\Psi-x^2\Psi^{\dagger}\Psi\right)\ ,
\label{freefermion}
\ee
where the fermion satisfies the commutation relations
\be
\left\{\Psi^{\dagger}(x),\Psi(x')\right\}
=\delta(x-x')\ .
\ee
The vacuum -- the flat spacetime with the linear dilaton and the
tachyon condensate $T=-\mu xe^{2x}$ -- corresponds to the 
fermi sea, $h(x,p)\equiv\half(p^2-x^2)\le -\mu$, or equivalently   
$-\sqrt{x^2-2\mu}\equiv p_-\le p\le p_+\equiv\sqrt{x^2-2\mu}$.

It is possible to deform the fermi surface. To see it, let us consider
the Hamilton-Jacobi equation of this system. It is given by
\be
\del_t S=-h(x,\del_x S)
=\half\left(x^2-(\del_x S)^2\right)\ ,
\ee
where $S(x;t)$ is Hamilton's principal function and 
$p=\del_x S$. 
Taking the partial derivative of this equation w.r.t. $x$, 
we obtain the Euler equation \cite{Polchinski:1991uq}
\be
\del_tp=x-p\del_xp\ .
\ee
The static solution $p=p(x)$ corresponds to the vacuum as we commented
above, whereas the non-static solutions $p=p(x;t)$ deform the static
fermi surface $p_{\pm}=\pm\sqrt{x^2-2\mu}$. 

The bosonisation $\del_x\phi=\pi\Psi^{\dagger}\Psi$ leads to 
Das-Jevicki-Sakita's collective field description of the double scaled
MQM \cite{Jevicki:1979mb,Das:1990ka}:
\be
S_B=\int {dtdx\over\pi}\left[
{(\del_t\phi)^2\over 2\del_x\phi}
-{1\over 6}(\del_x\phi)^3
+\left(\half x^2-\mu\right)\del_x\phi\right]\ .
\label{cft}
\ee
This is a ($1+1$)-dimensional string field theory, 
and the solutions to the classical equation of motion,
\be
\del_t\left({\del_t\phi\over\del_x\phi}\right)
-\half\del_x\left({\del_t\phi\over\del_x\phi}\right)^2
-{1\over 2}\del_x(\del_x\phi)^2+x=0\ ,
\label{bosonicEOM}
\ee
describe various different backgrounds.
With the relations \cite{Polchinski:1991uq},
\be
\del_x\phi={p_+-p_-\over 2}\ ,\qquad
{\del_t\phi\over\del_x\phi}
=-{p_++p_-\over 2}\ ,
\label{id}
\ee
the bosonic equation of motion (\ref{bosonicEOM}) is 
indeed equivalent to the fermionic ones 
\be
\del_tp_{\pm}=x-p_{\pm}\del_xp_{\pm}\ .
\label{euler}
\ee
Hence the deformations of the fermi surface correspond to generating 
different backgrounds in the ($1+1$)-dimensional string field theory 
\cite{Polchinski:1991uq}. 

Of our particular interest 
among non-static solutions to the Euler equation (\ref{euler})  
is
\be
\left(\sinh(t+t_0)x-\cosh(t+t_0)p\right)
(\sinh(t-t_0)x-\cosh(t-t_0)p)
=-\mu\sinh(2t_0)\ ,
\label{bbtfermisurface}
\ee
found in \cite{Minic:1991rk,Karczmarek:2003pv}, 
where $t_0\ge 0$ is a constant 
(see also related works 
\cite{Alexandrov:2002fh,Alexandrov:2003uh,Alexandrov:2003ut,Karczmarek:2004ph,
Das:2004hw,Mukhopadhyay:2004ff,Ernebjerg:2004ut,
Takayanagi:2004yr,Das:2004aq,Alexandrov:2004cg}   
on other non-static backgrounds).\footnote{The non-static background 
(\ref{bbtfermisurface}) is an example of the spacetime with spacelike
boundaries. Thus it offers an interesting laboratory for 
understanding the spacetime with cosmological horizon 
\cite{Das:2004aq}.
We would like to thank S. Das for bringing our attention to
the paper \cite{Das:2004aq}.}  
Incidentally this solution can be generated 
from the static solution 
by a similarity transformation 
$(x,p)\mapsto U^{-1}(x,p)U$
where $U=e^{-{\alpha\over 2}w_2}$
with $t_0={1\over 4}\ln\tanh\alpha$ 
and $w_2=w_{02}-w_{20}$ together with a scaling of 
$\mu$ \cite{Das:2004hw}. 
The generators $w_{rs}$ 
of the $W_{\infty}$ algebra are given by
\be
w_{rs}=\half e^{(r-s)t}(x-p)^r(x+p)^s\ ,
\ee
which satisfy
\be
\{w_{rs},w_{r's'}\}_{PB}
=(rs'-sr')w_{r+r'-1,s+s'-1}\ .
\ee

We shall consider the bosonic string theory, so will restrict
ourselves to one side of the fermi sea in the region $x<0$. 
For large $-x$,\footnote{More precisely, 
\be
-x\gg\sqrt{{4\mu\cosh(t+t_0)\cosh(t-t_0)\over\sinh(2t_0)}}\ .
\ee
} 
it is easy to find from the relations (\ref{id}) that
\be
\del_x\phi=-\mu x^{-1}
+{1\over 2}\left(\tanh(t+t_0)-\tanh(t-t_0)\right)x
+{\cal O}(x^{-3})\ .
\ee
The world-sheet tachyon $T$ is related to the collective field $\phi$,
for $x^{-2}\phi\ll 1$, via the dictionary \cite{Polchinski:1991uq}
\be
T+\half= e^{2q}\phi+{\cal O}(e^{4q})\ ,
\label{dictionary}
\ee
up to the leg pole factor which is not relevant in our case. 
Here we have defined $x\equiv -e^{-q}$.  
Then we have
\cite{Minic:1991rk,Karczmarek:2003pv}
\be
T=\mu qe^{2q}+{1\over 4}
\left({\sinh(2t_0)\over\cosh(t+t_0)\cosh(t-t_0)}-2\right)
\sim \mu qe^{2q}-\half\lambda\left(e^{2t}+e^{-2t}\right)\ ,
\label{bbtcorrespondence}
\ee
where we have defined $\lambda_{MM}=e^{-2t_0}\le 1$. 
The last expression is valid 
for $-|\ln\lambda_{MM}|\ll t\ll |\ln\lambda_{MM}|$ 
only for which $x^{-2}\phi\ll 1$ and the dictionary works.
Thus the non-static solution (\ref{bbtfermisurface}) describes the
bulk bouncing tachyon in the linear dilaton background 
\cite{Minic:1991rk,Karczmarek:2003pv}.
In particular, the last expression in (\ref{bbtcorrespondence})
corresponds to the world-sheet theory of the direct product of two
CFTs, the $c_L=25$ ($b=1$) Liouville and the $c_L=1$ ($b=i$) limit of
the Liouville theory (\ref{liouvilleaction}) 
with the dual Liouville potential turned on.

The non-static solution (\ref{bbtfermisurface}) suggests the following 
interpretation of the bulk bouncing tachyon 
\cite{Minic:1991rk,Karczmarek:2003pv}:
We will set $\mu$ to zero for simplicity. But the following picture 
is essentially the same for finite $\mu$. In the infinite past 
$t=-\infty$, the upper $p_+$ and lower half $p_-$ of the fermi surface
degenerate, that is, $p_{\pm}=-x$. This means that the fermi sea, 
$p_-\le p\le p_+$, is drained out.
There is nothing to be excited from the empty fermi sea -- no string
propagates. 
Since the dictionary (\ref{dictionary}) breaks down 
in particular at $t=-\infty$, 
we do not know it for sure but it is tempting to consider that the
tachyon is at $T=\infty$ at this point. 
Strings cannot propagate in the infinitely stiff tachyon media in
spacetime. 
As time goes by, the fermi sea is being filled until $t=0$ 
at which $p_{\pm}=\pm\tanh(t_0)x$ and it then starts draining.
It keeps on draining and is again emptied out in the infinite future
$t=+\infty$ at which $p_{\pm}=x$.

Rephrasing the process of the bulk bouncing tachyon, the spacetime (or
universe) starts out from nothing in the infinite past, is then
created and ``expands'' until $t=0$,   
and then ``contracts'' until it eventually dies off to
nothing in the infinite future. 
Note that there is a parallel between this picture and the full
S-brane of D-brane decay.
In the latter case, an unstable D-brane is created and then
annihilated, instead of the spacetime (or universe) 
\cite{Lambert:2003zr}.

Note also that when $t_0=0$ thus $\lambda_{MM}=1$, the upper and lower
half of the fermi surface always degenerate, that is, 
$p_{\pm}=\tanh(t)x$. Thus the fermi sea is drained out at all time.     
This point $\lambda_{MM}=1$ is similar to $\lambda_o=\half$ for the
full S-brane at which the system is thought to be at the bottom
of the tachyon potential. 

%%%%%%%%%%%%%%%%%%%%%%%%%%%%%%%%%%%%%%%%%%%%%%%%%%%%%%%%%%%%%%%%%
\subsection{The spectrum}
%%%%%%%%%%%%%%%%%%%%%%%%%%%%%%%%%%%%%%%%%%%%%%%%%%%%%%%%%%%%%%%%%
We would now like to understand the closed string spectrum in the
background (\ref{bbtfermisurface}). 
We will thus expand the collective field action (\ref{cft}) 
about the background $\phi_0$ as
\be
\phi(t,x)=\phi_0(t,x)
+\sqrt{\pi}\eta(t,x)\ ,
\ee
and look at the quadratic part w.r.t. the fluctuation $\eta$.
The quadratic action is given by
\be
S_{2}=\half\int {dtdx\over\del_x\phi_0}
\left[\left(\del_t\eta-{\del_t\phi_0\over\del_x\phi_0}
\del_x\eta\right)^2
-(\del_x\phi_0)^2(\del_x\eta)^2\right]\ .
\ee
Following \cite{Alexandrov:2003uh}, we introduce a new coordinate
system $(\tau,y)$ by
\be
y=\phi_0(t,x)\qquad
\mbox{and}\qquad
\tau=t\ .
\ee
Then the quadratic action simplifies to the diagonal form
\be
S_{2}=\half\int d\tau dy
\left[{1\over(\del_x\phi_0)^2}\left(\del_{\tau}\eta\right)^2
-(\del_x\phi_0)^2(\del_y\eta)^2\right]\ .
\ee
With this form, the fluctuation spectrum can be found by solving the
equation of motion
\be
\left[\del_{\tau}\left({1\over(\del_x\phi_0)^2}\del_{\tau}\right)
-\del_y((\del_x\phi_0)^2\del_y)\right]\eta(\tau,y)=0\ .
\ee
In our application, we have the explicit form
\be
y\sim -{\sinh(2t_0)\over 4\cosh(t+t_0)\cosh(t-t_0)}x^2
\quad\mbox{for}\ ,\quad 
-x\sqrt{{\sinh(2t_0)\over4\mu\cosh(t+t_0)\cosh(t-t_0)}}
\gg 1\ .
\ee
One can then easily find that the equation of motion 
takes the form
\be
\left[\del_{\chi}^2-\del_{\ln(-y)}^2\right]\eta(\chi,\ln(-y))\sim 0\ ,
\qquad\mbox{for}\qquad 
-y\gg\mu\ ,
\label{etaeom}
\ee
where we have introduced a new time coordinate $\chi$ by
\be
e^{\chi}={\cosh(\tau+t_0)\over\cosh(\tau-t_0)}
\label{chi}
\ee
In fact the coordinate system $(\chi,\ln(-y))$ for $-y\gg\mu$
is an example of the
flat coordinate in \cite{Alexandrov:2003uh}.
The quadratic action becomes that in the flat metric 
in terms of $(\chi,\ln(-y))$.

Note that since the r.h.s. of (\ref{chi}) increases monotonically as a
function of $\tau$,  
the new time coordinate $\chi$ has a finite range,
\be
-2t_0 \le \chi \le 2t_0\ ,
\label{finiterange}
\ee
Note also that 
\be
\chi\sim 2\tau\ , \quad\mbox{for}\quad
-t_0\ll \tau\ll t_0\ .
\label{timerelation}
\ee
The fluctuation equation of motion (\ref{etaeom})
reduces to 
\be
\left[\del_t^2-\del_q^2\right]\eta(t,q)\sim 0\qquad
\mbox{for}\qquad
-t_0\ll t\ll t_0\quad\mbox{and}\quad -q\gg 1\ ,
\label{etaws}
\ee
where the dictionary (\ref{bbtcorrespondence}) is valid.
This is the region far away from the Liouville wall and 
the confining walls of the bouncing tachyon, where the theory
is approximately free.

Since the time $\chi$ has the finite range, we need to impose the
boundary conditions at $\chi=\pm 2t_0$. 
It seems natural to impose the Dirichlet boundary condition, 
\be
\eta(\chi=\pm 2t_0,\ln(-y))=0\ .
\ee
Then we have
\be
\eta\propto e^{i\omega\chi}-e^{4it_0\omega}e^{-i\omega\chi}
=e^{i\omega\chi}-\lambda_{MM}^{2i\omega}e^{-i\omega\chi}\ ,
\label{wavefunction}
\ee
with the closed string energy $\omega$ being quantised by
\be
\omega={\pi n\over 4t_0}
={\pi n\over 2\ln\lambda_{MM}}\ ,
\qquad\qquad
n\in\mathbb{Z}_+\ .
\label{mmeq}
\ee
Since the free wave equation (\ref{etaeom}) takes the form
(\ref{etaws}) with the relation (\ref{timerelation}) 
in the weak field approximation,
the energy $\omega$ here is the same as the one in the world-sheet
theory (\ref{liouvilleaction}).
The same remark below (\ref{eqTShG}) is applied regarding $n$ being in
$\mathbb{Z}_+$.

Note first that the closed string wavefunction (\ref{wavefunction})
precisely agrees with that of the massless limit of the TShG model,
discussed in section 3, where the reflection amplitude is given by
$-((\lambda_{TShG})_{ren})^{2i\omega}$.  
This agreement may suggest the Dirichlet boundary condition to be the
right choice.
Secondly we find the exact agreement between the energy
quantisation (\ref{mmeq}) in the matrix model and the one
(\ref{eqTShG}) in the TShG model.  

Incidentally the bulk rolling tachyon can be obtained by first
shifting $\tau\to\tau+\tau_0$ and $\chi\to\chi+2\tau_0$, 
and then taking the limit 
\be
\lambda_{MM}\to 0\ ,\quad
\tau_0\to -\infty\ ,\quad
\mbox{keeping}\quad
\kappa\equiv\lambda_{MM}e^{-2\tau_0}
\quad\mbox{fixed finite}\ ,
\ee
which corresponds to the time-dependent part of the tachyon condensate 
to be $T=-\half\kappa e^{-2t}$.
From (\ref{chi}), one can see that 
\be
\ln\kappa\le\chi\ .
\label{halfline}
\ee
Thus the time direction is a half-line in the $(\chi,\ln(-y))$
coordinate. Imposing the Dirichlet boundary condition at
$\chi=\ln\kappa$ yields
\be
\eta\propto e^{i\omega\chi}-\kappa^{2i\omega}e^{-i\omega\chi}\ ,
\ee
with $\omega\in\mathbb{R}_+$, which is in agreement with the
reflection amplitude in the $c_L=1$ timelike Liouville theory 
\cite{Strominger:2003fn,Schomerus:2003vv}.

The flat coordinate $(\chi,\ln(-y))$ is particularly a
``good'' frame at least for $-y\gg\mu$. To see it, let us look at the
interaction part of the expansion of the action. One can find that it
is given by 
\be
S_{int}\sim\int {d\chi d\ln(-y)\over\pi}
\left[-{1\over 6}e^{-\ln(-y)}(\sqrt{\pi}\del_{\ln(-y)}\eta)^3
+\half(\sqrt{\pi}\del_{\chi}\eta)^2
\sum_{n=1}^{\infty}e^{-n\ln(-y)}
(-\sqrt{\pi}\del_{\ln(-y)}\eta)^n\right]\ .
\label{int}
\ee
Thus the interaction becomes time-independent, essentially the same as
that of the static background, as in the case of
\cite{Karczmarek:2004ph} 
(see also a related discussion \cite{Ernebjerg:2004ut}). 
In particular the perturbative treatment seems to make sense as long
as $-y\gg\mu$ even near $\chi\sim\pm 2t_0$ (thus $|t|\gg t_0$), 
which we believe validates our result.

Finally another point to make is that the flat coordinate
$(\chi,\ln(-y))$ manifests the matrix cosmology picture of 
\cite{Karczmarek:2003pv}. The spacetime is (conformally) flat, 
but the time interval
in which the universe exists varies with the strength 
$\lambda_{MM}=e^{-2t_0}$ of the bulk bouncing tachyon as 
$-|\ln\lambda_{MM}|\le \chi\le |\ln\lambda_{MM}|$.
At $\lambda_{MM}=1$ the universe does not exist at all, which is
analogous to $\lambda_o=\half$ for the full S-brane
of D-brane decay. 
In fact the energy quantisation (\ref{mmeq}) indicates that there is
no excitation to be allowed at $\lambda_{MM}=1$. 

%%%%%%%%%%%%%%%%%%%%%%%%%%%%%%%%%%%%%%%%%%%%%%%%%%%%%%%%%%%%%%%%%%
\section{Timelike $c_L=1$ Liouville}
\setcounter{equation}{0}
%%%%%%%%%%%%%%%%%%%%%%%%%%%%%%%%%%%%%%%%%%%%%%%%%%%%%%%%%%%%%%%%%%

We now try to understand the energy quantisation
directly from the construction of the $c_L=1$ (thus $b=i$) limit of  
the timelike Liouville theory with the dual Liouville potential
turned on.
Our attempt will end incomplete. 
We will, however, be able to show how the restriction on the spectrum
comes about and 
that the energy quantisation we found in the previous sections could
be consistent with that restriction.

On this score, Teschner's construction \cite{Teschner:1995yf} of 2 and
3-pt functions is particularly useful. 
In short, this construction provides two different (dual) ways to
obtain the 2 and 3-pt functions via the recursion relations. 
Two results then have to match up.
This constraint could lead to the restriction on the spectrum.
The condition from the 2-pt function consistency 
will be given in (\ref{cosmoduals}) 
and is in fact similar to the ones (\ref{smomq}) 
and (\ref{tshmom}) of the (T)ShG model.
In the case of $b=i$, it is very easy to see that 
(\ref{bimomq}) (the $b=i$ limit of (\ref{cosmoduals})) could take 
exactly of the same form as  
(\ref{TShGq}) if an appropriate choice of the function
$f_{i/2}(\alpha|R)$ is made.  
We will now discuss it in more detail.

Teschner's construction is to consider auxiliary ($n+1$)-pt functions 
with one insertion of the $(2,1)$ or $(1,2)$ degenerate 
field in order to calculate $n$-pt functions. 
The $(2,1)$ and $(1,2)$ degenerate fields are given by
$\psi_{2,1}\equiv V_{-b/2}$ and $\psi_{1,2}\equiv V_{-1/2b}$
respectively, where we have defined 
\be
V_{\alpha}=e^{2\alpha\phi}\ ,
\ee
with the conformal dimension $\Delta_{\alpha}=\alpha(Q-\alpha)$.
The operator product expansions (OPEs) of any primary field with
$\psi_{2,1}$ and $\psi_{1,2}$ involve only two conformal families
\cite{Belavin:1984vu},  
\bea
V_{\alpha}V_{-b/2}&=&C^{\alpha-b/2}_{\alpha,-b/2}
\left[V_{\alpha-b/2}\right]
+C^{\alpha+b/2}_{\alpha,-b/2}
\left[V_{\alpha+b/2}\right]\ ,
\label{21ope}\\
V_{\alpha}V_{-1/2b}&=&C^{\alpha-1/2b}_{\alpha,-1/2b}
\left[V_{\alpha-1/2b}\right]
+C^{\alpha+1/2b}_{\alpha,-1/2b}
\left[V_{\alpha+1/2b}\right]\ .
\label{12ope}
\eea
The structure constant $C^{\alpha+b/2}_{\alpha,-b/2}$ is given
explicitly by \cite{Fateev:2000ik}
\be
C^{\alpha+b/2}_{\alpha,-b/2}=-\pi\mu\frac{\gamma(2b\alpha-1-b^2)}
{\gamma(-b^2)\gamma(2b\alpha)}\ ,
\label{strconst}
\ee
where $\gamma(x)=\Gamma(x)/\Gamma(1-x)$, and
$C^{\alpha-b/2}_{\alpha,-b/2}$ and 
$C^{\alpha-1/2b}_{\alpha,-1/2b}$ are set to be unity by the choice of
normalization.

A remark is in order; As we will show the details in Appendix A, 
the formula (\ref{strconst}) for the structure constant is invalid at
$b=i$. 
This is because there are an infinite number of contributions from the
screening integrals at the point $b=i$, since the combination 
$\int d^2z e^{2b\phi(z)}\int d^2w e^{2\phi(w)/b}$ 
of two screening integrals has zero net
charge. Summing them up then yields
\be
C^{\alpha+i/2}_{\alpha,-i/2}
=
\lim_{b\to i}\frac{\pi\mu\gamma(b^2)}
{\left(1-(\pi\mu\gamma(b^2))(\pi\widetilde{\mu}
\gamma(1/b^2))\right)^2}\ .
\ee
The precise prescription of taking the $b\to i$ limit was given by
Schomerus \cite{Schomerus:2003vv} (and 
Fredenhagen and Schomerus \cite{Fredenhagen:2004cj}) 
in their construction of 3-pt functions; setting  
$b^2=-1+i\epsilon$ and then sending $\epsilon\to 0_+$, 
in order to approach the limit from the
well-defined regime of the 3-pt functions. Then we have
\bea
\lim_{b\to i}\gamma(b^2)
&=&\lim_{\epsilon\to 0_+}\gamma(-1+i\epsilon)
=\lim_{\epsilon\to 0_+}\left({i\over\epsilon}\right)\ ,\nn\\
\lim_{b\to i}\gamma(1/b^2)
&=&\lim_{\epsilon\to 0_+}\gamma(-1-i\epsilon)
=\lim_{\epsilon\to 0_+}\left(-{i\over\epsilon}\right)\ .\nn
\eea
As we motivated ourselves in the TShG argument above, we would like to
consider $\mu,\widetilde{\mu}<0$. In order to have a finite result, we
renormalise the CCs, as in Strominger and
Takayanagi \cite{Strominger:2003fn},
by setting $\mu=\mu_{ren}\epsilon$ and 
$\widetilde{\mu}=\widetilde{\mu}_{ren}\epsilon$. Then the structure
constant in the $b\to i$ limit is given by
\be
C^{\alpha+i/2}_{\alpha,-i/2}
=\frac{i\pi\mu_{ren}}
{\left(1-(\pi\mu_{ren})(\pi\widetilde{\mu}_{ren})\right)^2}\ ,
\label{bistrconst}
\ee
where $\mu_{ren},\widetilde{\mu}_{ren}<0$.

Similarly the dual structure constant $C^{\alpha+1/2b}_{\alpha,-1/2b}$
can be calculated as \cite{Fateev:2000ik}
\be
C^{\alpha+1/2b}_{\alpha,-1/2b}=-\pi\widetilde{\mu}
\frac{\gamma(2\alpha/b-1-1/b^2)}
{\gamma(-1/b^2)\gamma(2\alpha/b)}\ ,
\label{dualstrconst}
\ee%
for $b\ne i$, and 
\be
C^{\alpha-i/2}_{\alpha,i/2}
=\frac{-i\pi\widetilde{\mu}_{ren}}
{\left(1-(\pi\mu_{ren})(\pi\widetilde{\mu}_{ren})\right)^2}\ ,
\label{bidualstrconst}
\ee
for $b=i$.
They can be obtained by replacing $b$ by $1/b$ and $\mu$ by
$\widetilde{\mu}$ in the expression of
$C^{\alpha+b/2}_{\alpha,-b/2}$. 

%%%%%%%%%%%%%%%%%%%%%%%%%%%%%%%%%%%%%%%%%%%%%%%%%%%%%%%%%%%%%%%%%%%%%%
\subsection{Two-Point Function}
%%%%%%%%%%%%%%%%%%%%%%%%%%%%%%%%%%%%%%%%%%%%%%%%%%%%%%%%%%%%%%%%%%%%%

The two-point function in Liouville theory takes the form  
\be
|z|^{2\Delta_{\alpha_1}}\langle 
V_{\alpha_1}(z)V_{\alpha_2}(0)\rangle
=\delta(\alpha_1+\alpha_2-Q)
+D(\alpha_1)\delta(\alpha_1-\alpha_2)\ .
\label{2pt}
\ee
The first term corresponds to 
$|z|^{2\Delta_{\alpha}}\langle 
V_{\alpha}(z)V_{Q-\alpha}(0)\rangle=1$
which defines the normalisation, 
$C^{\alpha-b/2}_{\alpha,-b/2}=C^{\alpha-1/2b}_{\alpha,-1/2b}=1$,
that we adopted above.

We now consider an auxiliary 3-pt function with an insertion of the
(2,1) degenerate field,
\be
G_3(z_1,z_2,z;\alpha_1,\alpha_2)\equiv
\langle V_{\alpha_1}(z_1)V_{\alpha_2+b/2}(z_2)
V_{-b/2}(z)\rangle\ .
\label{aux3pt}\ee
The associativity and locality require that 
two ways of taking the OPEs (1) $z\to z_1$ 
and (2) $z\to z_2$ in (\ref{aux3pt}) give the same result, 
which in this case yields \cite{Fateev:2000ik} 
\be
D(\alpha)=C^{\alpha+b/2}_{\alpha,-b/2}D(\alpha+b/2)\ . 
\label{2ptshifteq}
\ee
This can be solved to
\be
D(\alpha)=f_{b/2}(\alpha|R)
\left(\pi\mu\gamma(b^2)\right)^{(Q-2\alpha)/b}
\frac{\gamma(2b\alpha-b^2)}{b^2\gamma(2-2\alpha/b+1/b^2)}\ ,
\label{2pt21}
\ee
where the function $f_{\nu}(\alpha|R)$ is a two-parameter
($\nu$ and $R$) family of periodic functions of $\alpha$ with 
period $\nu$. The parameter $R$ is 
defined by
$R=\ln\left[
\left(\pi\widetilde{\mu}\gamma(1/b^2)\right)^{b}/
\left(\pi\mu\gamma(b^2)\right)^{1/b}\right]$
and is an ``interval'' 
between two Liouville walls.\footnote{This is because formally
the Liouville wall is defined by $\mu e^{2b\phi_0}\sim 1$, and 
the dual Liouville wall by $\widetilde{\mu} e^{2\phi_0/b}\sim 1$.}
By a scaling argument, the periodic function
$f_{\nu}(\alpha|R)$ can depend on $\mu$ and $\widetilde{\mu}$
only through $R$.
In the $b=i$ case, as we discussed above, we have to deal with it
separately. However the result is as simple as 
replacing $\pi\mu\gamma(b^2)$ by (\ref{bistrconst}) and otherwise
setting $b=i$ in this expression.

A similar consideration by using the $(1,2)$ degenerate field 
$V_{-1/2b}$ yields 
\be
D(\alpha)=g_{1/2b}(\alpha|R)
\left(\pi\widetilde{\mu}\gamma(1/b^2)\right)^{b(Q-2\alpha)}
\frac{b^2\gamma(2\alpha/b-1/b^2)}{\gamma(2-2b\alpha+b^2)}\ ,
\label{2pt12}
\ee
The function $g_{\nu}(\alpha|R)$ is defined similarly as
$f_{\nu}(\alpha|R)$.  
Again, in the $b=i$ case, we replace $\pi\widetilde{\mu}\gamma(1/b^2)$
by (\ref{bidualstrconst}).

The periodic functions $f_{\nu}$ and $g_{\nu}$ are not
independent but related by  
the $\mathbb{Z}_2$-symmetry in (\ref{liouvilleaction}) 
under $b\leftrightarrow 1/b$ together with   
$\mu\leftrightarrow\widetilde{\mu}$. 
It is easy to show, for instance, by a heuristic manipulation of the
path integral expression of $D(\alpha)$, that 
\be
f_{1/2b}(\alpha|R)=g_{1/2b}(\alpha|-\!\! R)\ .
\ee

We have obtained the 2-pt function $D(\alpha)$ in two different ways, 
one (\ref{2pt21}) by using the $(2,1)$ degenerate field 
and the other (\ref{2pt12}) by the $(1,2)$
degenerate field. They must give the same result.
The second $\gamma$-function factors are shown to be the same.
Thus we have the constraint  
\be
f_{b/2}(\alpha|R)=e^{R(Q-2\alpha)}f_{1/2b}(\alpha|-\!\! R)\ .
\label{cosmoduals}
\ee

Note also that there is another $\mathbb{Z}_2$-symmetry under
$b\leftrightarrow -b$ together with $\alpha\leftrightarrow -\alpha$ 
and $\phi\leftrightarrow -\phi$, which leads to a constraint
\be
f_{-\nu}(-\alpha|-\!R)=f_{\nu}(\alpha|R)\ ,
\label{Z2}
\ee
where $\nu=b/2$ or $1/2b$. This can be used when we discuss the
$b=i$ case later.

There are two ways to think about the consistency condition
(\ref{cosmoduals});  
(1) it can be regarded as a condition on the periodic functions
$f_{b/2}(\alpha|R)$ and $f_{1/2b}(\alpha|-\!R)$, 
should the (\ref{cosmoduals}) hold for arbitrary momentum $\alpha$. 
(2) if we are to give up the first possibility, it may be
thought of as the restriction on the allowed momentum $\alpha$
for given periodic functions $f_{b/2}(\alpha|R)$ and
$f_{1/2b}(\alpha|-\!R)$.  

In the first case, one can see that the eq.(\ref{cosmoduals}) implies  
\bea
f_{b/2}(\alpha+1/2b|R)&=&e^{-R/b}f_{b/2}(\alpha|R)\ ,
\label{2ndperiod1}\\
f_{1/2b}(\alpha+b/2|-\! R)&=&e^{bR}f_{1/2b}(\alpha|-\! R)\ .
\label{2ndperiod2}
\eea
Thus $f_{b/2}(\alpha|R)$ and $f_{1/2b}(\alpha|-\! R)$ must be quasi-doubly
periodic. 
When the ratio of two periods is not real, there actually exist an
infinite number of such functions which can be constructed from the
Jacobi Theta functions. 
The ratio, however, is $b^2$, and thus this
being not real corresponds to the strongly coupled gravity regime
$1<c_L<25$.
It is a very interesting problem to understand this regime, but we
will not go into this case in the present paper.

In the case of $R=0$, two periodic functions
$f_{b/2}(\alpha|R)$ and $f_{1/2b}(\alpha|-\!R)$ are doubly periodic, 
and indeed the same from (\ref{cosmoduals}). 
Then it is well-known that, 
when the ratio of two periods $b^2$ is real and irrational, 
only such a function is constant. 
As noted in \cite{Teschner:1995yf},
although $b^2$ is rational for the much studied case of
two-dimensional quantum gravity coupling to the minimal models, 
it still provides a supportive argument -- if one assumes the
continuity w.r.t. $b$ -- for uniqueness of the solution
proposed in \cite{Dorn:1994xn,Zamolodchikov:1995aa}. 

We would like to make a comment here. Teschner's construction, as one
can see from the above discussion, requires the presence of the dual
Liouville potential. 
It may be interpreted as that the dual Liouville potential is generated
quantum mechanically \cite{Teschner:2001rv}.
The standard construction \cite{Zamolodchikov:1995aa} of the Liouville
theory further requires the dual CC to be at the
self-dual point $R=0$. This may be a manifestation of the  
$b\leftrightarrow 1/b$ duality of the 2 and 3-pt functions 
of \cite{Dorn:1994xn,Zamolodchikov:1995aa}, 
as noted in \cite{Fateev:2000ik}.
However, the canonical quantisation \cite{Teschner:2003en}
of the Liouville theory suggests that the Liouville field satisfies 
quantum mechanically the equation of motion without the dual Liouville
potential. 
Thus it seems not too clear whether or not the standard Liouville
theory should be thought of as the model with the dual Liouville
potential tuned at the self-dual point.

%%%%%%%%%%%%%%%%%%%%%%%%%%%%%%%%%%%%%%%%%%%%%%%%%%%%%%%%%%%%%%%%
\subsubsection{The $b=i$ case}
%%%%%%%%%%%%%%%%%%%%%%%%%%%%%%%%%%%%%%%%%%%%%%%%%%%%%%%%%%%%%%%%

Since our primary interest is in the $b=i$ case, we shall restrict
ourselves to this case.
Two periods $b/2$ and $1/2b$ become minus of each other when $b=i$, 
so they are degenerate. 
If the constraints (\ref{2ndperiod1}) and (\ref{2ndperiod2}) 
were to be satisfied for arbitrary $\alpha$, it would 
impose the factor $e^{iR}$ to be $1$. 
However it is impossible except for $R=0$,
since $R$ is pure imaginary,  
if one assumes $\mu$ and $\widetilde{\mu}$ to be real that is the
case of physical interest. 

Thus for generic $R$
the only possibility is to regard the consistency condition 
(\ref{cosmoduals}) as the restriction on the momentum $\alpha$.

The constraints (\ref{Z2}) from the $\mathbb{Z}_2$-symmetry, under
$b\leftrightarrow -b$ together with $\alpha\leftrightarrow -\alpha$ 
and $\phi\leftrightarrow -\phi$,
then relates two periodic functions by
\be
f_{-i/2}(\alpha|-\!R)=f_{i/2}(-\alpha|R)\ .
\label{z2rel}
\ee
Thus the restriction (\ref{cosmoduals}) on the momentum $\alpha$ 
now becomes 
\be
\frac{f_{i/2}(\alpha|R)}{f_{i/2}(-\alpha|R)}
=e^{-4i\alpha\ln\lambda_{lcft}}\ ,
\label{bimomq}
\ee
where 
\be
\ln\lambda_{lcft}\equiv -{iR\over 2}
=\ln\left(
\frac{\lambda_{ren}}
{(1-\lambda_{ren}^2)^2}
\right)\ ,
\qquad\mbox{and}\qquad\lambda_{ren}
\equiv\pi\sqrt{\mu_{ren}\widetilde{\mu}_{ren}}\ .
\ee
Notice its similarity to (\ref{TShGq}).
It is clear that if $f_{i/2}(\alpha|R)$ is an even function,
the constraint (\ref{bimomq}) is exactly of the same form as
(\ref{TShGq}). 

We will restrict ourselves to the case 
$\omega\equiv\alpha\in\mathbb{R}$ 
which corresponds to the timelike theory. 
Then the constraint (\ref{bimomq}) can be rewritten as
\be
4\omega\ln\lambda_{lcft}=2\pi n
+i\ln\left(
{f_{i/2}(\omega|R)\over f_{i/2}(-\omega|R)}
\right)\ ,
\qquad n\in\mathbb{Z}\ .
\label{engq}
\ee

From the CFT viewpoint, the choice of the periodic function
$f_{i/2}(\alpha|R)$, however, is rather arbitrary at this stage. 
If the function $f_{i/2}(\omega|R)$ is even or  
(quasi-)doubly periodic, that is, 
$f_{i/2}(\omega|R)=e^{2\pi ik}f_{i/2}(\omega+\pi/4\ln\lambda_{lcft}|R)$,  
where $k\in \mathbb{Z}$ and $f_{i/2}(0|R)<\infty$,
and $f_{i/2}(\omega+i/2|R)=f_{i/2}(\omega|R)$,
the constraint (\ref{engq}) reduces to   
the energy quantisation (\ref{eqTShG}) and (\ref{mmeq}). 
We also note that the TShG and matrix model result
(\ref{wavefunction}) may suggest $f_{i/2}(\omega|R)$ to be simply $1$.
Then by the field renormalisation 
$V_{\omega}\to
(e^{\pi i}\widetilde{\mu}_{ren}/\mu_{ren})^{i\omega/2}V_{\omega}$ 
the reflection amplitude becomes 
$D(\omega)=-\lambda_{lcft}^{2i\omega}$ in agreement with the results
in the previous sections.

%%%%%%%%%%%%%%%%%%%%%%%%%%%%%%%%%%%%%%%%%%%%%%%%%%%%%%%%%%%%%%%%%%%%%%%
\subsection{Three-Point Function}
%%%%%%%%%%%%%%%%%%%%%%%%%%%%%%%%%%%%%%%%%%%%%%%%%%%%%%%%%%%%%%%%%%%%%%%
Next we repeat a similar analysis for the 3-pt functions. 
In this case, we consider an auxiliary 4-pt function with an insertion
of the (2,1) degenerate field,
\be
G_4(z_1,z_2,z_3,z;\alpha_1,\alpha_2,\alpha_3)\equiv
\langle V_{\alpha_1}(z_1)V_{\alpha_2}(z_2)
V_{\alpha_3}(z_3)V_{-b/2}(z)\rangle\ .
\label{aux4pt}\ee
We will leave the details to Teschner's original paper 
\cite{Teschner:1995yf}. 
The crossing symmetry of the 4-pt function leads to a recursion
relation (eq.(21) and (26) of \cite{Teschner:1995yf})
for the structure constants, similar to the one (\ref{2ptshifteq}) for
the 2-pt functions, 
which can be solved to 
\cite{Dorn:1994xn, Zamolodchikov:1995aa}
\bea
C(\alpha_1,\alpha_2,\alpha_3)&=&
F_b(\alpha_1,\alpha_2,\alpha_3|R)
\left(\pi\mu\gamma(b^2)b^{2-2b^2}\right)^{(Q-\sum_i\alpha_i)/b}
e^{I_Q(\alpha_1,\alpha_2,\alpha_3)}\ ,
\label{3ptfunc}
\eea
where $F_{\nu}(\alpha_1,\alpha_2,\alpha_3|R)$ is a two-parameter 
($\nu$ and $R$) family periodic function of $\alpha_i$s with 
period $\nu$. It is symmetric under the permutations of
$\alpha_i$s. 
In the special case when $\alpha_3\to 0$, it must reduce to 
$f_{\nu/2}(\alpha|R)$, where $\alpha=\alpha_1=\alpha_2$, 
for the 2-pt function;\footnote{When $\alpha_3$ is sent to zero, 
the structure constant $C(\alpha_1,\alpha_2,\alpha_3)$ vanishes unless
$\alpha_2=\alpha_1$ or $\alpha_2=Q-\alpha_1$. In the $b=i$ limit, 
however, this point is subtle and requires a slightly nontrivial
procedure \cite{Runkel:2001ng}.} 
$F_b(\alpha,\alpha,0|R)=f_{b/2}(\alpha|R)$.
The function $I_Q(\alpha_1,\alpha_2,\alpha_3)$ is defined in the
appendix.

A similar consideration with the $(1,2)$ degenerate field leads to
\bea
C(\alpha_1,\alpha_2,\alpha_3)\!\!\!&=&\!\!\!
F_{1/b}(\alpha_1,\alpha_2,\alpha_3|-\!R)
\left(\pi\widetilde{\mu}\gamma(1/b^2)b^{-2+2/b^2}
\right)^{b(Q-\sum_i\alpha_i)}
\!e^{I_Q(\alpha_1,\alpha_2,\alpha_3)}\ ,
\label{3ptfuncdual}
\eea
Two results (\ref{3ptfunc}) and (\ref{3ptfuncdual}) have to match up. 
It then requires that
\be
F_b(\alpha_1,\alpha_2,\alpha_3|R)
=e^{R(Q-\sum_{i=1}^3\alpha_i)}
F_{1/b}(\alpha_1,\alpha_2,\alpha_3|-\!R)\ ,
\label{momq3pt}
\ee
similar to the one (\ref{cosmoduals}) for the 2-pt functions.
Note that this indeed reduces to that for the 2-pt functions when
$\alpha_3$ is sent to zero.

%%%%%%%%%%%%%%%%%%%%%%%%%%%%%%%%%%%%%%%%%%%%%%%%%%%%%%%%%%%%%%%%%
\subsubsection{The $b=i$ case}
%%%%%%%%%%%%%%%%%%%%%%%%%%%%%%%%%%%%%%%%%%%%%%%%%%%%%%%%%%%%%%%%%

We now focus on the $b=i$ case. 
The (dual-)CC dependent factors in (\ref{3ptfunc})
and (\ref{3ptfuncdual}) are to be replaced by (\ref{bistrconst}) and 
(\ref{bidualstrconst}) respectively, as in the case of 2-pt
functions. 
The nontrivial functional part $e^{I_Q(\alpha_1,\alpha_2,\alpha_3)}$
in this limit was first found by Runkel and Watts 
\cite{Runkel:2001ng} as a $c=1$ limit of the minimal models for the
spacelike theory, and then obtained as a certain $c_L=1$ limit of the
Liouville theory by Schomerus and Fredenhagen
\cite{Schomerus:2003vv,Fredenhagen:2004cj}.
In particular the 3-pt functions for the timelike theory were proposed
in \cite{Schomerus:2003vv}. 
Since the explicit form of this
function is not relevant at the level of our discussion, we will leave
it for their original papers. 

Similarly to the constraint
(\ref{z2rel}) from the $\mathbb{Z}_2$-symmetry, we have
\be
F_{-i}(\alpha_1,\alpha_2,\alpha_3|-\!R)
=F_{i}(-\alpha_1,-\alpha_2,-\alpha_3|R)\ .
\ee
Thus the momentum restriction is given by
\be
\frac{F_i(\alpha_1,\alpha_2,\alpha_3|R)}
{F_i(-\alpha_1,-\alpha_2,-\alpha_3|R)}
=e^{-2i\sum_{i=1}^3\alpha_i\ln\lambda_{lcft}}
=(-1)^{\sum_{i=1}^3n_i}\ .
\label{bimomq3pt}
\ee
where we have plugged $\alpha_i=\omega_i=\pi n_i/2\ln\lambda_{lcft}$ 
in, in the last equality.
So for the constraint (\ref{engq}) for the 2-pt functions to agree
with the energy quantisation (\ref{eqTShG}) and (\ref{mmeq}), 
the l.h.s. cannot be equal to $-1$, since
otherwise it would have implied that 
$f_{i/2}(\alpha|R)=-f_{i/2}(-\alpha|R)$.
Hence the l.h.s. must be equal to $1$ 
for the agreement of the CFT construction with the TShG and matrix
model results. 
It is then required that
\be
\sum_{i=1}^3n_i=2n\ , \qquad n\in\mathbb{Z}_+\ .
\ee
This may simply imply the selection rule for the sum of energies in
the 3-pt correlations. 
As noted in the end of the last subsection, the choice 
$F_i(\alpha_1,\alpha_2,\alpha_3|R)=1$ may be preferred.

In sum, it can be consistent at this stage to have the discrete
spectrum 
\be
\omega = {\pi n\over 2\ln\lambda_{lcft}}
= {\pi n\over 2\ln\left(
\lambda_{ren}/(1-\lambda_{ren}^2)^2\right)}\ ,
\qquad n\in \mathbb{Z}_+\ ,
\label{energyq}
\ee
in constructing the $c_L=1$ (thus $b=i$) limit of the timelike
continuation of (\ref{liouvilleaction}).
The TShG and matrix model results seem to suggest that the
unspecified functions $f_{i/2}(\alpha|R)$ and 
$F_i(\alpha_1,\alpha_2,\alpha_3|R)$ may simply be equal to $1$ 
(together with a field renormalisation
$V_{\omega}\to
(e^{\pi i}\widetilde{\mu}_{ren}/\mu_{ren})^{i\omega/2}V_{\omega}$),
as noted before.

However, the most important check is the compatibility 
of our proposed spectrum with the
conformal bootstrap equations \cite{Belavin:1984vu}.
The conformal bootstrap might fix the possible form of the function
$F_i(\alpha_1,\alpha_2,\alpha_3|R)$ (so $f_{i/2}(\alpha|R)$), 
or in the worst case disprove any form to be inconsistent.
To test the conformal bootstrap is beyond the scope of our paper. 
It remains to be seen.

For the spacelike $c_L=1$ Liouville theory without the dual
Liouville potential (or depending on the interpretation, 
with the self-dual Liouville potential, that is the 
$R=0$ case in our notation), the conformal bootstrap was checked 
numerically by Runkel and Watts \cite{Runkel:2001ng}, 
whereas for the timelike $c_L=1$ Liouville theory of 
\cite{Schomerus:2003vv,Strominger:2003fn}, 
it is yet to be checked, since in particular the timelike theory
cannot be obtained by a naive analytic continuation as emphasised 
in \cite{Schomerus:2003vv}.

%%%%%%%%%%%%%%%%%%%%%%%%%%%%%%%%%%%%%%%%%%%%%%%%%%%%%%%%%%%%%%%%%%%
\subsection{Limiting Cases}
%%%%%%%%%%%%%%%%%%%%%%%%%%%%%%%%%%%%%%%%%%%%%%%%%%%%%%%%%%%%%%%%%%

As noted already, the energy quantisation (\ref{energyq}) mimics
exactly that of the system in a finite time interval $\Delta T$, where
\be
\Delta T=2\ln\lambda_{lcft}
=2\ln\left[{\lambda_{ren}\over (1-\lambda_{ren}^2)^2}
\right]\ .
\ee
In fact the interval $|\Delta T|$ is the distance between two
(renormalised) Liouville walls.

Here we would like to comment on two extreme cases, 
(1) $\Delta T\to \infty$ and
(2) $\Delta T\to 0$.
We are interested in the limiting spectrum and the (dual-)CC dependent
factor in the 2 and 3-pt functions. 
We will restrict ourselves to the range $0\le\lambda_{ren}\le 1$, 
since the renormalisation of the couplings (\ref{bistrconst}) and
(\ref{bidualstrconst}) were obtained in the form of geometric series
with the convergence radius $|\lambda_{ren}|<1$ as shown in the
appendix. 

In the case (1), there are two ways to reach the limit; 
$\lambda_{ren}\to 0$ and $\lambda_{ren}\to 1$. 
We take the $\lambda_{ren}\to 0$ limit by
\bea
\widetilde{\mu}_{ren}\to 0\ ,\quad n\to \infty\ ,&&\quad
\mbox{keeping}\quad\mu_{ren}\quad\mbox{and}\quad
\omega\sim{\pi n\over 2\ln\lambda_{ren}}
\quad\mbox{fixed finite}\ ,\\
\mu_{ren}\to 0\ ,\quad n\to \infty\ ,&&\quad
\mbox{keeping}\quad\widetilde{\mu}_{ren}\quad\mbox{and}\quad
\omega\sim{\pi n\over 2\ln\lambda_{ren}}
\quad\mbox{fixed finite}\ .
\eea
In both cases the spectrum becomes continuous, 
as the interval $|\Delta T|$ is now infinite. 
The (dual-)CC dependent factor 
for each case simply takes the form respectively of  
\be
(i\pi\mu_{ren})^{i\sum_{i=1}^3\omega_i}\ ,
\qquad\mbox{and}\qquad
(-i\pi\widetilde{\mu}_{ren})^{-i\sum_{i=1}^3\omega_i}\ .
\label{singlelimit}
\ee
This is indeed the result one would expect for the $c_L=1$ limit 
\cite{Strominger:2003fn,Schomerus:2003vv} of the timelike
Liouville theory with a single Liouville potential.\footnote{However,
  see the comment made right before the section 5.1.1.} 
These limits provide the ingredient Liouville theories in the TShG
model in section 3.
It also suggests the renormalisation of the TShG coupling
\be
(\lambda_{TShG})_{ren}=\lambda_{ren}\quad \longrightarrow\quad
\lambda_{lcft}={\lambda_{ren}\over(1-\lambda_{ren}^2)^2}\ ,
\ee
away from $|(\lambda_{TShG})_{ren}|\ll 1$.

We next take the $\lambda_{ren}\to 1$ limit 
(thus $\pi\mu_{ren}=1/\pi\widetilde{\mu}_{ren}$) by
\be
\lambda_{ren}\to 1\ ,\quad n\to \infty\ ,\quad
\mbox{keeping}\quad
\omega\sim -{\pi n\over 4\ln(1-\lambda_{ren}^2)}
\quad\mbox{fixed finite}\ .
\ee
Again the spectrum becomes continuous for the same reason,
and the CC dependent factor reduces to
\be
(-1)^{\half\sum_{i=1}^3n_i}
\times (i\pi\mu_{ren})^{i\sum_{i=1}^3\omega_i}\ .
\ee
The sign factor $(-1)^{\half\sum_{i=1}^3n_i}$ may be absorbed into a
field renormalisation. Thus this case is essentially equivalent to  
the $\lambda_{ren}\to 0$ cases.

In the case (2), we need to take the coupling $\lambda_{ren}$ 
to the critical value
\be
\lambda_c=\{\lambda\,|\,\lambda^2+\sqrt{\lambda}-1=0\ ;
0\le\lambda\le 1\}\ .
\ee
This corresponds to $\lambda_{MM}=1$ of the matrix model. 
The spectrum disappears, as the interval $|\Delta T|$ is now zero.  
As remarked before, this point is an analogue of 
$\lambda_o=\half$ for the full S-brane of D-brane decay.

However, we note that since $\Delta T=0$ is nothing but 
the self-dual point $R=0$, 
we have an alternative isolated solution to the consistency conditions 
(\ref{bimomq}) and (\ref{bimomq3pt}) -- the continuous
spectrum, which thus is essentially equivalent to the $c_L=1$ limit of  
\cite{Strominger:2003fn,Schomerus:2003vv}.  

%%%%%%%%%%%%%%%%%%%%%%%%%%%%%%%%%%%%%%%%%%%%%%%%%%%%%%%%%%%%%%%%%%
\section{Summary and discussions}
%%%%%%%%%%%%%%%%%%%%%%%%%%%%%%%%%%%%%%%%%%%%%%%%%%%%%%%%%%%%%%%%%

We have proposed that the closed string energy in the bulk bouncing
tachyon background is to be quantised 
as if strings were trapped in a finite time interval.
The massless limit of the TShG model provides us with the most
intuitive argument for why and how the energy should be quantised.
The dual matrix model of the ($1+1$)-dimensional string theory 
helps us to take the common speculation on the bulk tachyon
condensation more seriously, 
and make the analogy between the bulk bouncing tachyon and the full
S-brane of D-brane decay clearer. 
It has then proved to give a strong support for our proposal, 
manifesting the intuition of strings in a finite time interval in a
concise manner. 
Our attempt of constructing the $c_L=1$ limit of the Liouville theory
with the dual Liouville potential has ended short of completion.
It, however, has provided a different way to reach 
the energy quantisation from the CFT viewpoint, which can be
consistent with the results in the TShG and matrix models, 
at least at the level of our analysis. 
The crucial test is to check whether our proposed discrete energy
spectrum is compatible with the conformal bootstrap constraints.
This remains to be seen.
 
The bulk tachyon might not have as interesting applications nor
as much motivation to consider it as the open string tachyon. 
However, the matrix cosmology of the bulk bouncing tachyon 
offers an interesting toy model of the Big-Bang/Crunch cosmology.
It might be interesting to consider an application of the bulk
bouncing tachyon to cosmology in higher dimensions, 
though in which taking into account the
back-reaction from the string pair productions to the geometry 
would be a challenging problem.

Finally there is a certain parallel between the bulk bouncing
tachyon and the Schwarzshild bubble of \cite{Witten:1981gj}.
It might be worthwhile to revisit the question discussed in 
\cite{DeAlwis:2002kp}.

%%%%%%%%%%%%%%%%%%%%%%%%%%%%%%%%%%%%%%%%%%%%%%%%%%%%%%%%%%%%%%%%%%
\section*{Acknowledgement}
%%%%%%%%%%%%%%%%%%%%%%%%%%%%%%%%%%%%%%%%%%%%%%%%%%%%%%%%%%%%%%%%%%
The author would like to thank Alex Flournoy for enjoyable discussions 
and especially 
Oren Bergman for helpful discussions
and comments on the manuscript. This work is in part supported by the 
Israel Science Foundation under grant no. 101/01-1.

%%%%%%%%%%%%%%%%%%%%%%%%%%%%%%%%%%%%%%%%%%%%%%%%%%%%%%%%%%%%%%%%%%
\setcounter{section}{1}
\renewcommand{\theequation}{\Alph{section}.\arabic{equation}}
\setcounter{equation}{0}
\section*{Appendix A: Calculation of Structure Constant}
%%%%%%%%%%%%%%%%%%%%%%%%%%%%%%%%%%%%%%%%%%%%%%%%%%%%%%%%%%%%%%%%%%

In this appendix, we will show the details of the computation of the
structure constant $C^{\alpha+b/2}_{\alpha,-b/2}$, in particular, with
emphasis on the peculiarity of the $b=i$ case.

From the OPE (\ref{21ope}), the structure constant may be given by
\bea
C^{\alpha+b/2}_{\alpha,-b/2}
&=&\lim_{\stackrel{z_1\to\infty ,z_3\to 0}{z_2\to 1}}
|z_1|^{4\Delta_{Q-\alpha-b/2}}
\sum_{n,m=0}^{\infty}{(-\mu)^{n}(-\widetilde{\mu})^m\over n!m!}
\\
&\times&
\langle V_{Q-\alpha-b/2}(z_1)V_{\alpha}(z_2)
\left(\int d^2u e^{2b\phi(u)}\right)^n
\left(\int d^2v e^{2\phi(v)/b}\right)^m
V_{-b/2}(z_3)\rangle_{free}\ ,\nn
\eea
where $\langle\cdots\rangle_{free}$ denotes the correlation function
of free CFT.
Except for $b=i$, the only contribution comes from $(n,m)=(1,0)$ in
the sum, due to the ``charge'' conservation in the presence of the
background charge $-2Q$.  
Then we only need to evaluate the single integral, which gives us  
(\ref{strconst}).

However, in the case of $b=i$, since the combination
$\int d^2z e^{2b\phi(z)}\int d^2w e^{2\phi(w)/b}$ 
of two screening integrals has zero net charge, 
there is an infinite number of contributions from 
all the terms of $(n,m)=(k+1,k)$ with $k\in\mathbb{Z}_+\cap\{0\}$.
Therefore 
we now need to evaluate the Dotsenko-Fateev integral
\cite{Dotsenko:1984ad}
\bea
I_{m,n}\left(\{\alpha_i\};b\right)&\equiv&
\lim_{\stackrel{z_1\to\infty ,z_3\to 0}{z_2\to 1}}
|z_1|^{4\Delta_{\alpha_1}}
{(-\mu)^{n}(-\widetilde{\mu})^m\over n!m!}\nn\\
&&\hspace{1.5cm}\times
\int\langle\prod_{i=1}^3V_{\alpha_i}(z_i)
\prod_{k=1}^nV_b(u_k)\prod_{j=1}^mV_{1/b}(v_j)
\rangle_{free}\,d^2u_kd^2v_j\ ,
\eea
with $(n,m)=(k+1,k)$ with $k\in\mathbb{Z}_+\cap\{0\}$.

This integral can be read off from the Liouville 3-pt function, 
since the 3-pt function was so constructed that it
obeys 
\be
I_{n,m}(\alpha_i;b)=\mbox{Res}_{\stackrel{}{\sum_i\alpha_i=Q-nb-m/b}}
C(\alpha_1,\alpha_2,\alpha_3)\ ,
\ee
where the 3-pt function is given by
\bea
C(\alpha_1,\alpha_2,\alpha_3)&=&
\left(\pi\mu\gamma(b^2)b^{2-2b^2}\right)^{(Q-\sum_i\alpha_i)/b}
e^{I_Q(\alpha_1,\alpha_2,\alpha_3)}\ ,
\label{3ptfunction}
\eea
and we have defined
\be
e^{I_Q(\alpha_1,\alpha_2,\alpha_3)}\equiv
\frac{\Upsilon_0\Upsilon(2\alpha_1)\Upsilon(2\alpha_2)
\Upsilon(2\alpha_3)}
{\Upsilon(\sum_i\alpha_i-Q)\Upsilon(\alpha_1+\alpha_2-\alpha_3)
\Upsilon(\alpha_2+\alpha_3-\alpha_1)
\Upsilon(\alpha_3+\alpha_1-\alpha_2)}\ ,
\label{IQ}
\ee
and the replacement
\bea
\left.\left[\pi\mu\gamma(b^2)b^{2-2b^2}\right]^{(Q-\sum_i\alpha_i)/b}
\right|_{\sum_i\alpha_i=Q-nb-m/b}
\hspace{-.5cm}\rightarrow
\left[\pi\mu\gamma(b^2)b^{2-2b^2}\right]^n
\left[\pi\widetilde{\mu}\gamma(1/b^2)(1/b)^{2-2/b^2}\right]^m
\ ,\nn
\eea
is required in (\ref{3ptfunction}).
The special function $\Upsilon(x)$ is defined by
\be
\ln\Upsilon(x)\equiv
\int_0^{\infty}{dt\over t}\left[
(Q/2-x)^2e^{-2t}-\frac{\sinh^2(Q/2-x)t}{\sinh(bt)\sinh(t/b)}
\right]\ ,
\ee
and 
\be
\Upsilon_0\equiv \left.\frac{d\Upsilon(x)}{dx}
\right|_{x=0}\ .
\ee
It has zeroes at $x=-nb-m/b$ and 
$x=Q+nb+m/b$ with $n, m\in \mathbb{Z}_+\cap \{0\}$.
Also it is clearly invariant under $b\leftrightarrow 1/b$.

By making use of the formula
\bea
\Upsilon(x-nb-m/b)&=&(-1)^{nm}b^{-n(n+1)b^2}
b^{m(m+1)/b^2}\left(b^{-1+2bx}\right)^n
\left(b^{1-2x/b}\right)^m\\
&&\hspace{-2cm}\times\left[
\prod_{l=1}^{n}\prod_{k=1}^{m}
(x-lb-k/b)^{-2}
\prod_{l'=1}^{n}\gamma(bx-l'b^2)
\prod_{k'=1}^{m}\gamma(x/b-k'/b^2)
\right]^{-1}\Upsilon(x)\ ,\nn
\eea
one can find that 
\bea
I_{n,m}(\{\alpha_i\};b)&=&
b^{4(n-m)}
\left[\pi\mu\gamma(b^2)\right]^n
\left[\pi\widetilde{\mu}\gamma(1/b^2)\right]^m
\nn\\
&&\times
\prod_{l=1}^{n}\prod_{k=1}^{m}(lb+k/b)^{-2}
\prod_{l'=1}^n\gamma(-l'b^2)
\prod_{k'=1}^m\gamma(-k'/b^2)\label{DFintegral}\\
&&\hspace{-3cm}\times
\prod_{i=1}^3\Biggl[
\prod_{l''=0}^{n-1}\prod_{k''=0}^{m-1}
(2\alpha_i+l''b+k''/b)^{-2}
\prod_{l'''=0}^{n-1}\gamma(-2b\alpha_i+1-l'''b^2)
\prod_{k'''=0}^{m-1}\gamma(-2\alpha_i/b+1-k'''/b^2)
\Biggr]
\ .\nn
\eea

\noindent
We now apply this formula to the $b=i$ case by taking the following
limit 
\be
C^{\alpha+i/2}_{\alpha,-i/2}
=\lim_{\epsilon\to 0_+}\lim_{\eta\to 0}\,\sum_{n=0}^{\infty}
I_{n+1,n}(Q-\alpha-b/2,\alpha,-b/2+\eta;b)\Biggr|_{b^2=-1+i\epsilon}\ ,
\ee
where we first take the $\eta\to 0$ limit and then $\epsilon\to 0_+$.
Without this limiting procedure, the Dotsenko-Fateev integral
(\ref{DFintegral}) is not well-defined.
Note that the result crucially depends on how to perform the
regularisation and take the limit. 
Our reasoning for choosing this particular limit is 
the following; 
There is a zero and pole at $\alpha=-b/2$ in (\ref{DFintegral})
irrespective of the value of $b$, 
so we first regularise them by introducing the regulator $\eta$ for
arbitrary $b$. 
Then since there appear additional zeroes and poles when 
$b^2\to -1$, we further introduce the regulator $\epsilon$. 
A particular choice $b^2=-1+i\epsilon$ is instructed from Schomerus's
construction of the $c_L=1$ Liouville 3-pt functions 
\cite{Schomerus:2003vv}.  

Using repeatedly the formulae
\bea
&&
\gamma(x-n)=(-1)^n\prod_{l=1}^n(x-l)^{-2}\gamma(x)\ ,
\qquad
\gamma(x+n)=(-1)^n\prod_{l=0}^{n-1}(x+l)^2\gamma(x)\ ,\nn\\
&&\gamma(x)\gamma(1-x)=1\ ,
\qquad \gamma(x)\gamma(-x)=-{1\over x^2}\ ,
\eea
one can find that
\be
C^{\alpha+i/2}_{\alpha,-i/2}=i\pi\mu_{ren}\sum_{n=0}^{\infty}
(n+1)(\pi\mu_{ren})^n(\pi\widetilde{\mu}_{ren})^n
=\frac{-i\pi\mu_{ren}}{(1-(\pi\mu_{ren})(\pi\widetilde{\mu}_{ren}))^2}\ ,
\label{Cminus}
\ee
where we have renormalized the CCs by taking
the limits $\mu,\,\widetilde{\mu}\to 0$ 
and $\epsilon\to 0_+$ with $b^2=-1+i\epsilon$, 
while $\mu_{ren}\equiv -i\mu\gamma(b^2)$ and 
$\widetilde{\mu}_{ren}\equiv i\widetilde{\mu}\gamma(b^2)$ 
kept fixed, as in Section 3.
Since the series in (\ref{Cminus}) is only convergent for 
$\lambda_{ren}^2=(\pi\mu_{ren})(\pi\widetilde{\mu}_{ren})<1$, 
either the theory with $|\lambda_{ren}|>1$ is ill-defined, 
or one might take the final expression of (\ref{Cminus}) to define the
theory with $|\lambda_{ren}|>1$
by means of analytic continuation.

%%%%%%%%%%%%%%%%%%%%%%%%%%%%%%%%%%%%%%%%%%%%%%%%%%%%%%%%%%%%%%%%%
%%%%%%%%%%%%%%%%%%%%%%%%%%%%%%%%%%%%%%%%%%%%%%%%%%%%%%%%%%%%%%%%%%%

\end{document}